\documentclass{article}
\usepackage{spconf,amsmath,graphicx}
\usepackage{xcolor}
\usepackage{amsmath,bbm}
\usepackage{amssymb}
\usepackage{booktabs}
\usepackage{multirow}
\usepackage{cite}
\usepackage[hidelinks]{hyperref}
\usepackage[mathscr]{euscript}
\usepackage{fancyhdr}
\usepackage{indentfirst}

\usepackage{caption}
\usepackage{subcaption}
\usepackage[nodisplayskipstretch]{setspace}
\usepackage{flushend}

\usepackage{fancyhdr}
\fancypagestyle{firstpage}{
  \fancyhf{}
  \setlength{\headheight}{20mm}
  \renewcommand{\topmargin}{-15mm}
  \fancyhead[C]{\vspace{5mm} \small \textit{Proc. IEEE ICIP 2023}, Kuala Lumpur, Malaysia, Oct. 2023\\ © 2023 IEEE. Personal use of this material is permitted. Permission from IEEE must be obtained for all other uses, in any current or future media, including reprinting/republishing this material for advertising or promotional purposes, creating new collective works, for resale or redistribution to servers or lists, or reuse of any copyrighted component of this work in other works.}}

\AtBeginDocument{%
  \addtolength\abovedisplayskip{-0.5\baselineskip}%
  \addtolength\belowdisplayskip{-0.5\baselineskip}%
}

\DeclareMathAlphabet\mathbfcal{OMS}{cmsy}{b}{n}

\title{Base Layer Efficiency in Scalable Human-Machine Coding}
\name{Yalda Foroutan, Alon Harell, Anderson de Andrade, and Ivan V. Baji\'{c}\thanks{This work was supported by NSERC and Intel Labs.}}
\address{School of Engineering Science, Simon Fraser University, Burnaby, BC, Canada}
\begin{document}
%
\maketitle


%
\begin{abstract}
A basic premise in scalable human-machine coding is that the base layer is intended for automated machine analysis and is therefore more compressible than the same content would be for human viewing. Use cases for such coding include video surveillance and traffic monitoring, where the majority of the content will never be seen by humans. Therefore, base layer efficiency is of paramount importance because the system would most frequently operate at the base-layer rate. In this paper, we analyze the coding efficiency of the base layer in a state-of-the-art scalable human-machine image codec, and show that it can be improved. In particular, we demonstrate that gains of 20-40\% in BD-Rate compared to the currently best results on object detection and instance segmentation are possible. 
\end{abstract}


\vspace{-0.3cm}
\section{Introduction}
\vspace{-0.15cm}

\thispagestyle{firstpage}

\label{sec:intro}
Traditionally, image and video codecs have been developed to minimize the bitrate required to support human viewing of the visual content. Codecs such as JPEG, JPEG2000, and H.26X have become enabling technologies for many multimedia services. Increasingly, however, visual content is also ``viewed'' by machines for the purpose of automated analysis. Examples include traffic monitoring, visual surveillance, autonomous driving, and others. This has spurred interest in the development of codecs optimized for machine-based analysis~\cite{Hyomin_ICASSP_2018,dfc_for_collab_object_detection}, which have demonstrated significant bit savings compared to coding for human viewing. Reasons for these bit savings have also been supported by rate-distortion theory~\cite{choi2022scalable}.

In applications such as traffic monitoring, analysis tasks -- such as object detection, tracking, speed estimation -- are supposed to run continuously, while human viewing may be necessary on occasion to assess a certain situation of interest, for example an accident. For these applications, codecs would need to support both machine analysis and human viewing. This has inspired human-machine scalable codecs~\cite{scalable_face_TMM_2021,yan2021sssic,choi2022scalable}, where the base layer supports machine analysis tasks, while the enhancement layer supports human viewing. 

Our focus in this paper is on a scalable human-machine image  codec~\cite{choi2022scalable}, which presents state-of-the-art (SOTA) results for the cases of object detection + human viewing and object detection + instance segmentation + human viewing. 
In applications that require such scalable coding, the efficiency of the base layer is crucial, because the machine analysis runs continuously while human viewing is only needed occasionally. Our main goal here is to improve the efficiency of the base layer. Our contributions are as follows: \vspace{-5pt}
\begin{itemize}
    \item We show that, despite its good performance, the base layer in~\cite{choi2022scalable} is inefficient. We explain the source of this inefficiency, and propose a way to optimize the base-layer efficiency in human-machine scalable coding. \vspace{-5pt}
    \item We demonstrate coding gains of up to 20-40\% in BD-Rate over~\cite{choi2022scalable} for the base tasks of object detection and instance segmentation, setting, to our knowledge, new SOTA rate-accuracy results on these tasks. \vspace{-5pt}
\end{itemize}

Section~\ref{sec:related_work} briefly reviews related work. The proposed method is described in Section~\ref{sec:proposed}. Experimental results are presented in Section~\ref{sec:experiments}, followed by conclusions in Section~\ref{sec:conclusion}.

\begin{figure}
    \centering
    \includegraphics[width=0.6\linewidth]{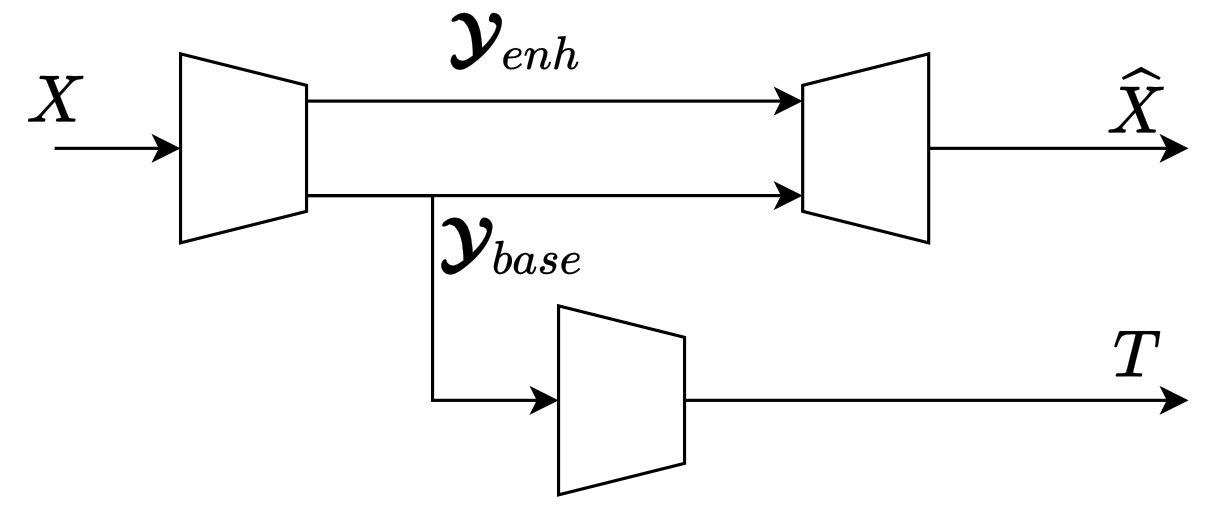}
    \vspace{-0.3cm}
    \caption{Scalable human-machine coding by latent-space partitioning~\cite{choi2022scalable}}
    \label{fig:hyomin_scalable}
    \vspace{-5pt}
\end{figure}

\vspace{-0.3cm}
\section{Related work}
\label{sec:related_work}
\vspace{-0.15cm}
Although coding for machines can be performed using conventional handcrafted codecs~\cite{Hyomin_ICASSP_2018,dfc_for_collab_object_detection,JPEG_PRL23}, learning-based codecs offer greater flexibility to exploit redundancies encountered in this kind of compression. Learning-based image codecs have come a long way recently~\cite{toderici-variable,minnen2018joint,balle2018variational,cheng2020learned, hu2020coarse,xie2021enhanced}, rivaling the performance of the best handcrafted codecs, especially on perceptual metrics like Structural Similarity Index Metric (SSIM). The general processing chain of recent learned image codecs can be described as $X \overset{g_a}{\to} \mathbfcal{Y} \to \widehat {\mathbfcal{Y}} \overset{g_s}{\to} \widehat{X}$, where $X$ is the input image, $g_a$ is a learnable analysis transform, $\mathbfcal{Y}$ is the latent-space representation of the input image, which is quantized to $\widehat {\mathbfcal{Y}}$, $g_s$ is a learnable synthesis transform, and $\widehat{X}$ is the approximation to the input image. Upon deployment, and during our evaluation, the quantized latent representation $\widehat {\mathbfcal{Y}}$ is entropy-coded and decoded. Such codecs are trained end-to-end using a Lagrangian loss function: 

\begin{equation}
    \mathcal{L} = R + \lambda \cdot D,
    \label{eq:loss}
\end{equation}
where $R$ is the total rate estimate for 
$\widehat {\mathbfcal{Y}}$ and any side information needed for its encoding and decoding, 
and $D$ is the distortion between $X$ and  $\widehat{X}$. The scalar $\lambda$ 
can be adjusted to achieve the desired balance between compression efficiency and reconstructed image quality.

Recently, the concept of latent-space scalability~\cite{LSS_ICIP2021,choi2022scalable} was proposed for scalable human-machine coding. Here, the latent representation is partitioned into base and enhancement portions, $\mathbfcal{Y} = \{\mathbfcal{Y}_{base}, \mathbfcal{Y}_{enh}\}$, such that $\mathbfcal{Y}_{base}$ feeds the machine analysis task $T$, while both $\mathbfcal{Y}_{base}$ and $\mathbfcal{Y}_{enh}$ are used for input reconstruction, as shown in Fig.~\ref{fig:hyomin_scalable}. Such a system can be trained end-to-end using a generic loss function:
\begin{equation}
    \mathcal{L} = R_{base} + R_{enh} + \lambda_{base} \cdot D_{base} + \lambda_{enh} \cdot D_{enh},
    \label{eq:scalable_loss}
\end{equation}
where $R_i$ are the rate estimates for layer $i\in\{base,enh\}$, including any side information, and $D_i$ are distortions related to the specific layer. In~\cite{choi2022scalable}, two systems were demonstrated based on this concept: a 2-layer system whose base layer supports object detection using YOLOv3~\cite{redmon2018yolov3}, and a 3-layer system whose base layer supports object detection using Faster R-CNN~\cite{ren2016faster}, together with the second base (or first enhancement) layer that supports instance segmentation using Mask R-CNN~\cite{he2017mask}. In each case, state-of-the-art rate-accuracy results were demonstrated on the machine task, while remaining comparable to the state-of-the-art image codecs for human viewing as the enhancement task. However, despite its good performance, the base layer of~\cite{choi2022scalable} turns out to be inefficient. In the next section, we explain the source of this inefficiency and present a way to improve it.

\vspace{-0.3cm}

\section{Proposed methods}
\label{sec:proposed}

\vspace{-0.15cm}

The idea behind using the loss function~(\ref{eq:scalable_loss}) to optimize the codec in~\cite{choi2022scalable} was that the base task-relevant information will be steered towards $\mathbfcal{Y}_{base}$ because distortion $D_{base}$ depends only on $\mathbfcal{Y}_{base}$ and not $\mathbfcal{Y}_{enh}$. While this is true, it is only part of the story. To appreciate what happens during training, consider the following simplified\footnote{This is a simplified scenario because, in reality, features are being created and placed into $\mathbfcal{Y}_{base}$ or $\mathbfcal{Y}_{enh}$ simultaneously, whereas here, for simplicity, we assume that features are created first, then their placement is decided.} scenario: imagine that a feature $f$ has already been created and the goal of training is to decide whether to place it in $\mathbfcal{Y}_{base}$ or $\mathbfcal{Y}_{enh}$ based on the loss function~(\ref{eq:scalable_loss}). Consider two extreme cases:

\textbf{Case 1}: feature $f$ carries base task-relevant information; note that such a feature also carries information relevant to the enhancement task of input reconstruction.  If such a feature is placed into $\mathbfcal{Y}_{base}$, then $R_{base}$ will increase and both $D_{base}$ and $D_{enh}$ will decrease, as the feature will be used for both tasks. On the other hand, if such a feature is placed into $\mathbfcal{Y}_{enh}$, then $R_{enh}$ will increase, but only $D_{enh}$  will decrease, because the feature will not be used for the base task. Hence, base task-relevant features are encouraged to end up in $\mathbfcal{Y}_{base}$.

\textbf{Case 2}: feature $f$ carries no base task-relevant information, only enhancement-relevant information. If such a feature is placed into $\mathbfcal{Y}_{base}$, then $R_{base}$ will increase and $D_{enh}$ will decrease, because both $\mathbfcal{Y}_{base}$ and $\mathbfcal{Y}_{enh}$ are used for the enhancement task. Similarly, if such a feature is placed into $\mathbfcal{Y}_{enh}$, then $R_{enh}$ will increase and $D_{enh}$ will decrease, because both $\mathbfcal{Y}_{base}$ and $\mathbfcal{Y}_{enh}$ are used for the enhancement task. Therefore, based on~(\ref{eq:scalable_loss}), there is no preference for placing $f$ in either $\mathbfcal{Y}_{base}$ or $\mathbfcal{Y}_{enh}$, and such a feature may end up anywhere in the latent space.
 
Based on the analysis above, we conclude that features that are not relevant to the base task (case 2) may still end up in $\mathbfcal{Y}_{base}$. In other words, the base layer of the codecs presented in~\cite{choi2022scalable} is likely inefficient, containing more information than strictly necessary for the base task. In the next section, we describe how base layer efficiency can be improved.    

\vspace{-0.3cm}

\subsection{Base layer}
\label{subsec:base}
\vspace{-0.15cm}

As our framework is based on~\cite{choi2022scalable}, we consider three machine analysis tasks as our base layer tasks: object detection using YOLOv3, object detection using Faster R-CNN, and instance segmentation using Mask R-CNN. Our results demonstrate that we can improve the rate-accuracy efficiency on these tasks by 20-40\% in BD-Rate. 

As noted in the previous section, the source of inefficiency in the base layer is the task-irrelevant information, which can end up in $\mathbfcal{Y}_{base}$ when base and enhancement layers are trained in parallel, as in~\cite{choi2022scalable}. For this reason, we employ sequential training, where the base layer is trained first, then frozen, followed by training of the enhancement layer. Conceptually, the base layer processing chain can be described as $X \overset{g_b}{\to} \mathbfcal{Y}_{base} \to \widehat {\mathbfcal{Y}}_{base} \overset{g_t}{\to} T$, where $g_b$ is the base analysis transform, and $g_t$ accomplishes the task $T$ from $\widehat{\mathbfcal{Y}}_{base}$. The key to making an efficient base layer is to realize an Information Bottleneck (IB)~\cite{IB_Allerton1999} for task $T$: 
\begin{equation}
    \min_{p(\widehat{
    y}_{base} | x)} \quad I(X; \widehat{\mathbfcal{Y}}_{base}) - \beta \cdot I(\widehat{\mathbfcal{Y}}_{base} ; T),
\label{eq:IB}
\end{equation}
where $I(\cdot;\cdot)$ is the mutual information~\cite{Cover_Thomas_2006}, $p(\widehat{
y}_{base} | x)$ is the mapping from the input image to the base-layer latent representation, and $\beta>0$ is the IB Lagrange multiplier~\cite{IB_Allerton1999}.

In our case, 
$p(\widehat{
y}_{base} | x)$ 
is $g_b$ followed by quantization. Hence, when input $X$ is given, $\widehat{\mathbfcal{Y}}_{base}$ is fully determined, so we have $I(X; \widehat{\mathbfcal{Y}}_{base}) = H(\widehat{\mathbfcal{Y}}_{base}) - H(\widehat{\mathbfcal{Y}}_{base} | X) = H(\widehat{\mathbfcal{Y}}_{base})$, because $H(\widehat{\mathbfcal{Y}}_{base} | X) = 0$. Here, $H(\cdot)$ and $H(\cdot | \cdot)$ are the entropy and conditional entropy~\cite{Cover_Thomas_2006}, respectively. Moreover, since decreasing  $-\beta\cdot I(\widehat{\mathbfcal{Y}}_{base} ; T)$ (i.e., increasing $I(\widehat{\mathbfcal{Y}}_{base} ; T)$) is supposed to improve the task accuracy, we take $\lambda_{base} \cdot D_{base}$ as the proxy for $-\beta\cdot I(\widehat{\mathbfcal{Y}}_{base} ; T)$. Therefore, in our case, the IB~(\ref{eq:IB}) for task $T$ becomes:
\begin{equation}
    \min_{g_b,g_t} \quad H( \widehat{\mathbfcal{Y}}_{base}) + \lambda_{base} \cdot D_{base},
\label{eq:IB_base}
\end{equation}
showing that it can be solved using a loss function analogous to~(\ref{eq:loss}). We employ~\cite{cheng2020learned} to realize $g_b$ and perform entropy estimation, and the Latent Space Transform (LST) from~\cite{choi2022scalable} to map $\widehat{\mathbfcal{Y}}_{base}$ to the task network features, where we use MSE as our distortion target during training. The number of channels in $\widehat{\mathbfcal{Y}}_{base}$ is set depending on the task, as described in the experiments. 

\vspace{-0.3cm}

\subsection{Enhancement layer}
\begin{figure}[t]
  \centering
  \includegraphics[width=\linewidth]{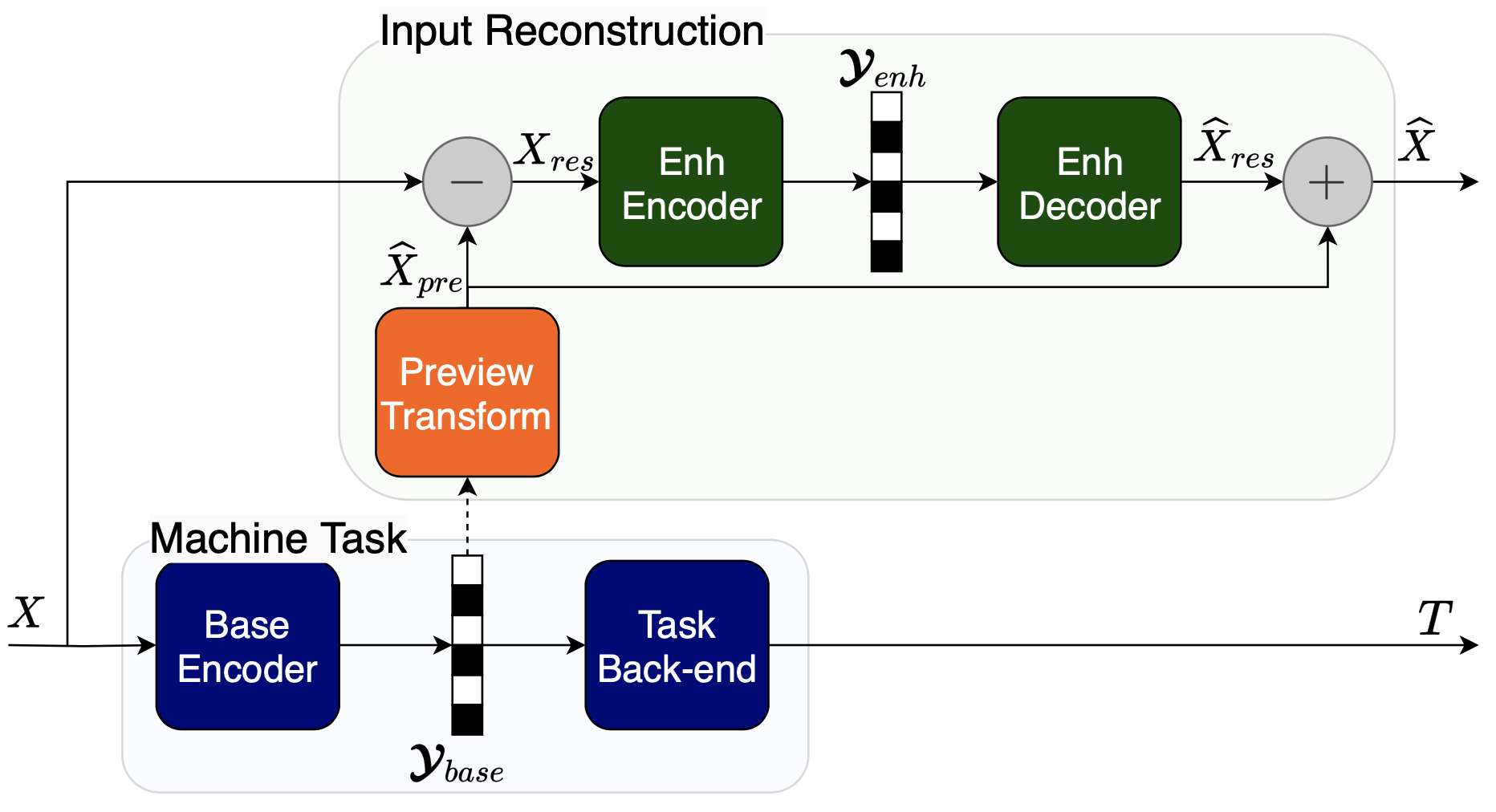}
  \vspace{-0.5cm}
  \caption{Sequentially-optimized two-task scalable codec: the base layer is optimized first, and then frozen while the enhancement layer is optimized.}
  \vspace{-5pt}
\label{figure:detailed_res}
\end{figure}

After training the base layer, we freeze it, and construct a ``preview transform'' that is meant to recover an approximation $\widehat X_{pre}$
of the input image $X$ from base features $\widehat{\mathbfcal{Y}}_{base}$, as shown in Fig.~\ref{figure:detailed_res}. This transform consists of a $1 \times 1$ convolutional layer to adjust the number of channels, followed by a synthesis transform $g_s$ from~\cite{cheng2020learned} to approximate the input image $X$. Next, we subtract the preview image $\widehat X_{pre}$ from the input $X$, resulting in a residual image $X_{res}$. This residual image is then encoded by~\cite{cheng2020learned}, which was fine-tuned for our setting using an MSE loss. Finally, we obtain the reconstructed input $\widehat X$ by adding $\widehat X_{pre}$ to the reconstructed residual $\widehat X_{res}$. The dashed line in Fig.~\ref{figure:detailed_res} indicates that during 
training of the enhancement layer, the base layer is frozen.

\vspace{-0.3cm}

\section{Experiments}
\label{sec:experiments}
\vspace{-0.2cm}

\subsection{Base layer}
 
As explained above, we begin by training and evaluating the base layer without the enhancement part. We do this using several tasks and models following~\cite{choi2022scalable}: object detection using YOLOv3~\cite{redmon2018yolov3} and Faster R-CNN~\cite{ren2016faster}, and instance segmentation using Mask R-CNN~\cite{he2017mask}. All base-layer experiments share the following two-stage training approach. Input images are random patches of size $256 \times 256$ from the JPEG-AI~\cite{JPEGAIDataset} and CLIC~\cite{CLIC} datasets in stage one and VIMEO-90K~\cite{Xue2019} in stage two, with a mini-batch size of 16. The first stage uses the Adam optimizer with a fixed learning rate of $10^{-4}$ for 500 epochs, followed by another 400 epochs with a polynomial decay of the learning rate every 10 epochs. The Lagrange multipliers as well as the number of base channels $L_{base}$ for each task are shown in Table~\ref{table:lambda}. 

\begin{table}[htbp]
    \caption{Task model parameters}
    \vspace{-5pt}
    \small
    \centering
    \setlength{\tabcolsep}{2pt} 
    \begin{tabular}{c| cccccc  c}
    \bottomrule
     Task model & \multicolumn{6}{c}{$\lambda_{base}$} & $L_{base}$ \\
    \hline
    Faster R-CNN  & 1.28e--5 & 3.2e--5 & 8e--5   & 2e--4  & 4e--4  & 5.5e--4 & 96 \\
    Mask R-CNN    & 1.28e--5 & 3.2e--5 & 8e-5   & 2e--4  & 4e--4  & 5.5e--4 & 128\\
    YOLOv3        & - &  5e--5   & 1e--4  & 4e--4  & 1e--3 & 2e--3 & 64\\
    \toprule
    \end{tabular}
    \label{table:lambda}
\end{table}

\begin{figure}[t]
  \centering
  \includegraphics[width=0.9\linewidth]{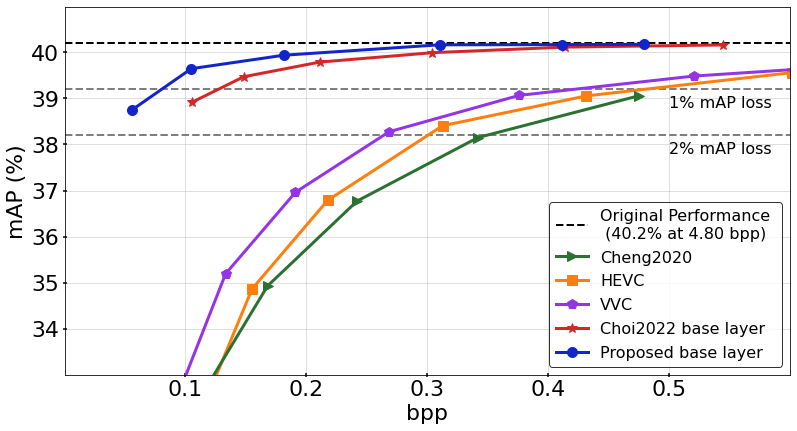}
  \vspace{-0.3cm}
  \caption{Evaluation of base layer for object detection task using Faster R-CNN backbone on COCO2017 validation set}
  \label{fig:obj}
\end{figure}

\begin{figure}[t]
  \centering
  \includegraphics[width=0.9\linewidth]{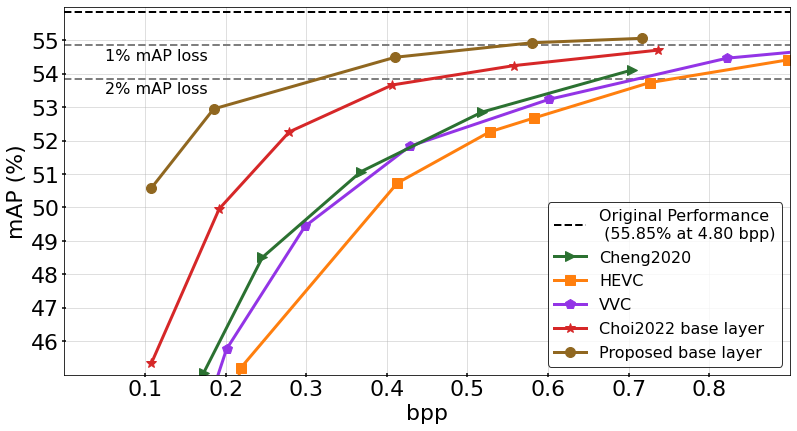}
  \vspace{-0.3cm}
  \caption{Evaluation of base layer for object detection task using YOLOv3 backbone on COCO2014 validation set}
  \label{fig:yolo}
\end{figure}

\begin{figure}[t]
  \centering
  \includegraphics[width=0.9\linewidth]{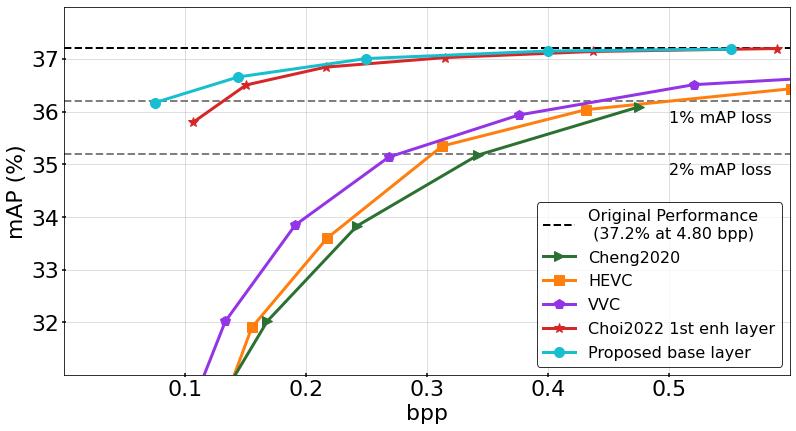}
  \vspace{-0.3cm}
  \caption{Evaluation of base layer for instance segmentation using Mask R-CNN backbone on COCO2017 validation set}
  \label{fig:seg}
\end{figure}

Task performance is then evaluated following the same procedure as in the 3-layer network described in~\cite{choi2022scalable}, using the 
COCO2017/COCO2014 validation set~\cite{EEEECOCO}, and Detectron2~\cite{wu2019detectron2}. We maintain the same formulation for the mean average precision (mAP) as in~\cite{choi2022scalable} and compare the performance of our proposed base layer with \cite{choi2022scalable} (which we refer to as Choi2022) alongside other benchmarks in Fig.~\ref{fig:obj}-\ref{fig:seg}.

Observing the three figures, it is evident that our base layer outperforms the previous SOTA by a significant margin. For example, Fig.~\ref{fig:obj}, shows that using Faster R-CNN, \cite{choi2022scalable} experiences a $1.3\%$ drop in mAP at $0.1$ bits per pixel (BPP), while our base layer only drops by $0.56\%$. Similarly, when using YOLOv3 
in Fig.~\ref{fig:yolo}, ~\cite{choi2022scalable} suffers a degradation of $1.15\%$ even at a relatively high bit-rate of over $0.7$ BPP, while our approach remains at less than $1\%$ reduction in mAP until approximately $0.5$ BPP. Lastly, we see that such improvement persist even in the more complex task of instance segmentation, seen in Fig.~\ref{fig:seg}.

To summarize the difference in performance between our proposed method and previous SOTA, we employ the Bj\o{}ntegaard Delta (BD) metric~\cite{Bjontegaard} for rate difference at equivalent 
accuracy (BD-Rate). We see significant savings in all three experiments, which are summarized in Table~\ref{table:base_bdrate}.

\begin{table}[tbp]
\vspace{10pt}
    \centering
    \captionsetup{justification=centering}
    \caption{BD-Rate of the proposed base layer relative to the base layer of~\cite{choi2022scalable}}
    \vspace{-5pt}
    \centering
    \begin{tabular}{c| cccccc c}
    \bottomrule
    Task model   & Faster R-CNN  & Mask R-CNN & YOLOv3\\[2pt]
    \hline
    \\[-9pt]
    BD-Rate (\%) &  --36.5       & --18.9     & --41.7\\ 
    \toprule
    \end{tabular}
    \label{table:base_bdrate}
    \vspace{-0.3cm}
\end{table}

\vspace{-0.4cm}
\subsection{Enhancement layer}

Once the base-layer training is complete, we proceed to 
train the enhancement layer. Because of the decoupling between the training of our base and enhancement layers, we can theoretically match various levels of image reconstruction to each of our base-layer models. However, in order to create a fair comparison with previous models, where base and enhancement performance are inherently linked, we match the relative qualities of our two layers.  
In training the enhancement layer, we freeze our 
base layer and initialize the 
residual using a pre-trained model~\cite{cheng2020learned} (Cheng2020 in figures) of varying quality index, provided by CompressAI~\cite{begaint2020compressai}. For the first three base qualities, we choose quality index 1, while for higher base qualities, we assign quality 3 from~\cite{cheng2020learned}. ‌It is worth noting that both quality levels contain 128 channels.

\begin{figure}[tbp]
  \centering
  \includegraphics[width=0.9\linewidth]{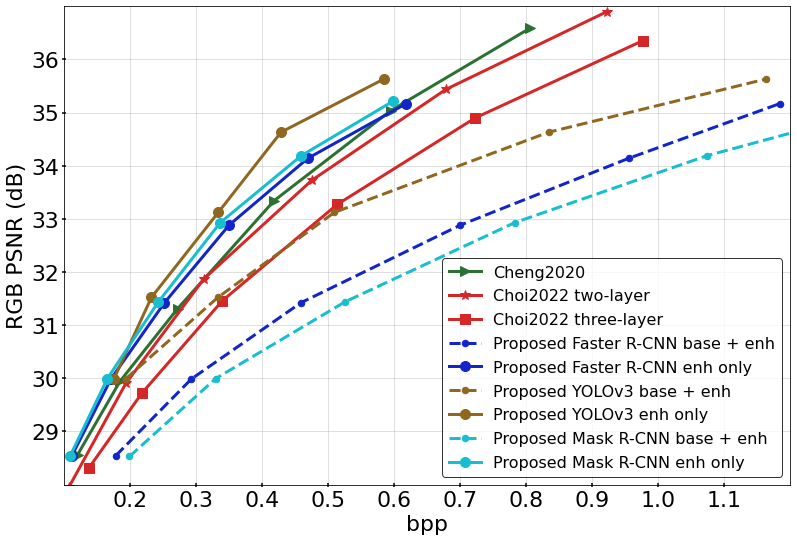}
  \vspace{-0.3cm}
  \caption{Comparison of input reconstruction performance using proposed base layers on Kodak dataset}
  \label{fig:res_res}
\end{figure}

The enhancement layer is trained for 300 epochs using random patches of $256 \times 256$ from the JPEG-AI~\cite{JPEGAIDataset} and CLIC~\cite{CLIC} datasets, with a mini-batch size of 16, the Adam optimization algorithm with a constant learning rate of $10^{-4}$ is used. The values of the $\lambda_{enh}$ are similar to the $\lambda_{enh}$ used in~\cite{choi2022scalable}. The performance of the enhancement layer is evaluated on Kodak~\cite{kodak} dataset, which consists of 24 uncompressed images. The results of the input reconstruction are shown in terms of PSNR vs. BPP in Fig.~\ref{fig:res_res}. Table~\ref{table:bdrate_enh} summarizes the BD-Rate results using Cheng2020 as the benchmark, including Choi2022 
and our proposed approach.

\begin{table}[tbp]  
    \caption{BD-Rate relative to \cite{cheng2020learned}}
    \vspace{-5pt}
    \centering
    \setlength{\tabcolsep}{6pt}
    \begin{tabular}{c| c |c |c} 
    \bottomrule
     Network & Setting & BPP & BD-Rate (\%) \\
     \hline
    \multirow{2}{*}{Choi2022} & 2-layer   & Total & 4.40   \\
                              & 3-layer & Total & 24.11 \\
    \hline
    \multirow{6}{*}{Proposed} & \multirow{2}{*}{Faster R-CNN} & Total & 71.46 \\ 
                              &                                          & Enh. only        & --7.41 \\

                              & \multirow{2}{*}{Mask R-CNN}   & Total & 92.86 \\ 
                              &                                          & Enh. only        & --10.92 \\
    
                              & \multirow{2}{*}{YOLOv3}       & Total & 28.90 \\ 
                              &                                          & Enh. only        & --16.95 \\
    
    \toprule
    \end{tabular}
    \label{table:bdrate_enh}
    \vspace{-0.2cm}
\end{table}  

As could perhaps have been expected, the improvement of our base layer came at the cost of some degradation to the enhancement-layer performance. We do see significant increase in bitrate when compared with 
Cheng2020, as well as the 
scalable approach of Choi2022~\cite{choi2022scalable}. Fortunately, however, when comparing our enhancement layer alone, we still see rate savings compared to re-transmitting the full image, with BD-Rate improvement (compared to~\cite{cheng2020learned}) ranging from $7-17\%$. In a practical setting, the choice between two approaches will depend on the relative frequency of the use of human vision compared to machine vision, denoted $f_T$. Using this, the relative rate of our approach and a human-vision encoder is given by:
\begin{equation}
    (1-f_T) \cdot\frac{R_{base}}{R} + f_T\cdot \frac{R_{base}+R_{enh}}{R},
    \label{eq:fraction}
\end{equation}
where $R$ is the bitrate of the single-layer encoder. Whenever the relative rate is smaller than 1, our approach will be preferable. Using BD-Rate estimates to approximate the ratios in Eq.~\ref{eq:fraction}, we see, that our models achieves overall rate savings so long as human viewing is used less than $59\%$, $34\%$, and $17\%$, for YOLOv3, Faster R-CNN, and Mask R-CNN, respectively. 

\vspace{-0.3cm}
\section{Conclusions}
\label{sec:conclusion}
\vspace{-0.2cm}

We have explained and demonstrated an inefficiency in the base layer in the previous state-of-the-art human-machine scalable image codec. To mitigate this problem, we proposed an improved base-layer training procedure achieving significant rate savings on multiple tasks. We then added a residual-based enhancement layer for input reconstruction. As expected, we saw some degradation in rate-distortion for the enhancement layer, which we believe can be reduced through further optimization of the enhancement layer. Nevertheless, in scenarios where machine analysis is needed more frequently than human viewing, our proposed method outperforms relevant single-layer learned image codecs.


\bibliographystyle{IEEEbib-abbrev} 
\small
\bibliography{ref}

\end{document}